# Thermodynamics of Polydomain Ferroelectric Bilayers and Graded Multilayers

Alexander L. Roytburd[1], and Julia Slutsker[2]

[1]Department of Materials Science and Engineering, University of Maryland, College Park MD 20742
[2]Materials Science and Engineering Laboratory, National Institute of Standards and Technology, Gaithersburg, MD 20899

The equilibrium domain structure and its evolution under an electric field in ferroelectric bilayers and graded multilayers are considered. The equilibrium bilayer is self-poled and contains a single-domain and a polydomain (with 180° domains) layers. The polarization of a graded multilayer proceeds by movement of wedge-like domains as a result of progressive transformation of polydomain layers to a single-domain state. The theory provides the principal explanation of peculiarities of dielectric behavior of graded ferroelectric films and can be applied to graded ferromagnetics and ferroelastics.

Unusual properties of graded ferroelectric films, particularly, polarization offset, and enhanced dielectric and pyroelectric responses have been a subject of extensive experimental and theoretical investigations for more than a decade [1, 2]. All theoretical explanations including Landau-Devonshire-Ginzburg phenomenological approach, transverse Ising models and first principles calculations considered the single-domain layers in bilayers or multilayers [2-12]. However, a domain structure in graded ferroelectric films and is effect on film properties remain unclear. It is especially timely to address this problem, since recent studies show that the domain structures appear even in very thin ferroelectric films [13-16]. In this paper we present the thermodynamic theory of a polydomain ferroelectric bilayer and a graded multilayer, which provides an important insight into a domain structure and its evolution in graded ferroic multilayers and films. The theory is based on the model of a graded multilayer as an ensemble of electrostatically interacting single-domain and polydomain layers. Each polydomain layer contains specific equilibrium domain fractions depending on the applied electric field. The multilayer is polarized through changing domain fractions and transition between the polydomain and single-domain states of the layers.

We start with a bilayer as the simplest model of graded ferroelectrics. The free energy density of a bilayer (or a multilayer consisting of a set of identical bilayers) under bias voltage between the top and bottom electrodes can be presented as follows [7]:

$$F = \alpha_1(\varphi_1(p_1) - EP_1) + \alpha_2(\varphi_2(p_2) - EP_2) + \frac{1}{2}\alpha_1\alpha_2 \frac{1}{\varepsilon_0}(P_1 - P_2)^2 + o(d/H) \quad , \quad (1)$$

where $\alpha_1 = h_1/H$ and $\alpha_2 = h_2/H$ are relative thicknesses of layers 1 and 2, and H is a thickness of a bilayer; $\varphi_1(p_1)$ and $\varphi_2(p_2)$ are the free energy densities of the layers as a function of the microscopic polarization $p_1$ and $p_2$. The spontaneous polarization $p_{01}$ and $p_{02}$ are solutions of the equations $\partial\varphi_1/\partial p_1 = 0$ and $\partial\varphi_2/\partial p_2 = 0$, where $p_{01} < p_{02}$ is considered. $P_1$ and $P_2$ are the macroscopical polarizations of the layers, which are equal to $p_1$ and $p_2$ for the single-domain layers or are dependent on fractions of 180° domains, $\beta_1$ and $\beta_2$, for the polydomain layers:

$$P_1 = (1 - 2\beta_1)p_1 \quad , \quad P_2 = (1 - 2\beta_2)p_2 \quad . \quad (2)$$

The direction of polarizations and an external field, E, are assumed to be normal to the layers, for simplicity. The first two terms in Eq.(1) correspond to the Gibbs free energies of the individual layers, while the third term describes the energy of electrostatic interactions between the layers ($\varepsilon_0$ is the dielectric constant of vacuum). All these terms are scale-independent, so they are the same for a bilayer and a periodic multilayer. The last term in Eq.(1) (d is a domain period) should include the scale-dependent contribution of domain walls and inhomogeneous electrostatic fields near the interfaces separating the polydomain layers. For d<H this term can be estimated as $1/\varepsilon_0 (p_{02} - p_{01})^2 \sqrt{l/H}$, and it can be omitted for not very thin layers, when H>>$l$ where $l$~1nm is a domain wall thickness [17]. The short-range interactions between the single-domain and polydomain layers can be neglected also for such not very thin films, while, as shown recently by many experimental and theoretical studies, electrostatic interactions play a dominant role [12-15].

The equilibrium of the bilayer is determined by the following set of equations $\frac{\partial F}{\partial \beta_1} = \frac{\partial F}{\partial \beta_2} = 0$, $\frac{\partial F}{\partial p_1} = \frac{\partial F}{\partial p_2} = 0$. In the absence of external field in a short circuited bilayer these equations have an infinite number of solutions: $p_1=p_{01}$, $p_2=p_{02}$, $P_1=P_2$, including the domain structure with zero net polarization, $\beta_1=\beta_2=1/2$ [18]. One of these solutions describes a bilayer consisting of a single-domain layer with polarization $p_1=p_{01}$ and a polydomain layer with a macroscopical polarization $P_2=(1-2\beta_2)p_{02}=p_{01}$, so an average polarization, which determines the bilayer surface charge, $P=\alpha_1 P_1+\alpha_2 P_2=p_{01}$ (Fig.1). All solutions correspond to a minimum of the free energy $F=\alpha_1\varphi_1(p_{01})+\alpha_2\varphi_2(p_{02})$. However, taking into consideration the energy of the domain walls and inhomogeneous fields, a polydomain/single-domain bilayer should be considered more stable than polydomain/polydomain one unless the electrodes are not perfect and create the depolarized field.

The polydomain/single-domain bilayer is the only solution of the equilibrium equations at E≠0, when only two of these equations remain to be compatible:

$$\frac{\partial F}{\partial p_1} = 0, \quad \beta_1 = 0, \quad \frac{d\varphi_1}{dp_1} = E - \frac{\alpha_2}{\varepsilon_0}(p_1 - P_2) = E + \frac{1}{\varepsilon_0}(P - P_1) \quad (3)$$

$$\frac{\partial F}{\partial \beta_2} = 0, \quad p_2 = p_{02}, \quad 0 = E + \frac{\alpha_1}{\varepsilon_0}(p_1 - P_2) = E + \frac{1}{\varepsilon_0}(P - P_2) \quad (4)$$

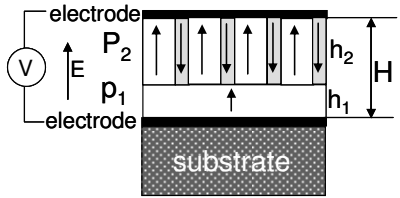

FIG.1 Ferroelectric bilayer consisting of a single-domain layer and a polydomain layer.

Eq.(4) shows that the equilibrium of polydomain layers requires zero net field, i.e. the external and internal fields should cancel each other. From the equations of states (3) and (4) the field dependence of the polarizations, the domain fraction and the displacement, D, follow:

$$P_2 = p_1 + \frac{\varepsilon_0}{\alpha_1}E = p_{02}(1-2\beta_2),$$

$$\beta_2 = \frac{1}{2}(1-(\frac{p_1}{p_{02}} + \frac{\varepsilon_0}{\alpha_1}\frac{E}{p_{02}})),$$

$$D = \alpha_1 P_1 + \alpha_2 P_2 + \varepsilon_0 E = p_1 + \frac{\varepsilon_0}{\alpha_1}E,$$ as well as the

equations determining $p_1$,

$$\frac{d\varphi_1}{dp_1} = \frac{1}{\alpha_1}E \quad \text{or}$$

$$\frac{d\varphi_1}{dp_1} \cong \frac{1}{\varepsilon_1 - \varepsilon_0}(p_1 - p_{01}) = \frac{1}{\alpha_1}E. \quad (5)$$

The equilibrium displacement, D(E) is presented in Fig.2. There are two singular points of D(E) dependence at critical fields $E_c$ and $E_{c1}$. $E_c$ is a coercive field of the lost of stability of the single-domain weak layer 1 in respect to switching of the direction of its polarization (see Eq.(5)) and, therefore, the total polarization of the bilayer switches. The domain fraction in layer 2 changes from $\beta_2 < 1/2$ to $\beta_2 > 1/2$. For free energy density,

$$\varphi_1(p_1) = \frac{1}{4(\varepsilon_1 - \varepsilon_0)}(-p_1^2 + \frac{p_1^4}{2p_{01}^2}),$$ which may include

effect of a substrate, $E_c = \frac{\alpha}{\varepsilon_1 - \varepsilon_0}\frac{1}{3\sqrt{3}}p_{01}$ ($\varepsilon_1$ is the dielectric constant of layer 1). For small $\alpha_1$, $E_c$ is much smaller than coercive field of a single layer film, so the thermodynamic hysteresis loop can be observed if $E_c < E_{c1}$.

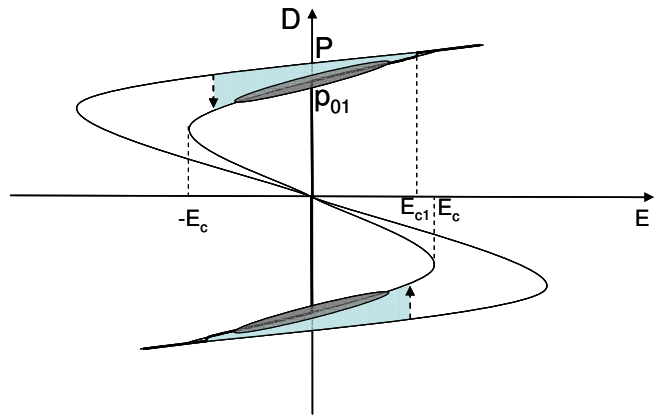

FIG.2. Equilibrium D(E) dependence. The stable equilibrium is shown by solid lines. The hystersis loops for E<$E_{c1}$ (dark areas) and for E≥$E_{c1}$ (light areas) are shown.

The other critical field, $E_{c1}$, corresponds to the transition of a polydomain state in layer 2 into a single-domain one ($\beta_2=0$). It can be estimated using linear Eq.(5)

$$E_{c1} = \frac{\varepsilon_1(p_{02} - p_{01})}{\alpha_1} \qquad (6)$$

Between these critical points the polarization of the polydomain layer, $P_2$, as well as the average polarization, $P$, are very close (with accuracy $\varepsilon_0/\varepsilon_1$) to the polarization of the single-domain layer $p_1$. The external field is concentrated in the single-domain layer increasing the low-field dielectric susceptibility of a bilayer by $1/\alpha$ times due to absence of the field in polydomain layers, i.e. the polydomain layer amplifies the dielectric response of the bilayer. Then,

$$D^{(1)} = \varepsilon^{(1)} E + p_{01}, \quad \varepsilon^{(1)} \equiv \varepsilon_1/\alpha_1, \quad E<E_{c1} \qquad (7)$$

The similar increase of dielectric susceptibility of a polydomain layer with dielectric dead layers [19] and the increase of mechanical compliance of layer composite containing layers with elastic domains was considered in [20]. In contrast with a dielectric dead layer [19] or an elastically passive layer in adaptive composite [17,20], in a ferroelectric bilayer a single-domain layer makes it poled and due to its strongly nonlinear stiffness, dramatically increases bilayer sensibility to electrical field and temperature change.

When the field exceeds $E_{c1}$ both layers are in single-domain states and the displacement becomes (in the linear approximation):

$$D^{(2)} = \varepsilon^{(2)} E + P_r^{(2)}, \quad \varepsilon^{(2)} = (\frac{\alpha_1}{\varepsilon_1} + \frac{\alpha_2}{\varepsilon_2})^{-1},$$
$$P_r^{(2)} = \varepsilon^{(2)}(\frac{\alpha_1}{\varepsilon_1} p_{01} + \frac{\alpha_2}{\varepsilon_2} p_{02}) \qquad (8)$$

Depending on the bilayer parameters and amplitude of an applied AC field, several different regimes of quasistatic polarization are possible. If $E_c<E<E_{c1}$, two identical hysteresis loops shifted up or down with respect to the axis $D=0$, depending on the initial self-poled states, should be observed. The widths of the loops are determined mainly by the dissipation of energy during evolution of the domain structure in layer 2.

When field increases polydomain layer 2 transforms to the single-domain state at $E_{c1}$. In Eq.(6) $E_{c1}$ is estimated neglecting nonlinearity of polarization in single-domain layer, $p_1$, as well as contribution of the domain structure to the energy of a polydomain layer. This effect is small at $d/H<1$ [20], however, when the field approaches $E_{c1}$ and $\beta_2$ approaches zero, the domain period grows, and the domain collapse leads to sharp increase of $D(E)$ [21, 22]. Since the nucleation of domains in layer 2 is needed for reverse polarization, the single-domain state in this layer can remain at field less than $E_{c1}$. It can lead to formation of asymmetric hysteresis loops (Fig. 2).

The $D(E)$ dependence of a ferroelectric bilayer reproduces the main peculiarities of polarization curves of graded films, such as up and down shifts of hysteresis loops and enhanced dielectric susceptibility [5,23,24]. The bilayer can be considered as the first two layers of a multilayer consisting of the sequence of layers with increasing spontaneous polarization, $p_{01}$, $p_{02}$ …$p_{0n}$ and with relative thicknesses, $\alpha_1,\alpha_2...\alpha_n$. In the absence of external field in the short circuited multilayer layer 1 with minimal spontaneous polarization is in a single-domain state, all other are in polydomain states. The minimum of free energy is reached if $p_1=P_1=P_2...P_n$ with $p_1=p_{01}$ and $P_i=p_{0i}(1-2\beta_i)$, $i=2…n$. Then, domain fractions at zero field are $\beta_i=1/2(1-p_{0i}/p_{01})$. To minimize interlayer energy 180° domains should form wedge-like macroscopical domains (Fig. 3) (the similar wedge-like domains are formed in bent polydomain ferroelastic films [25]).

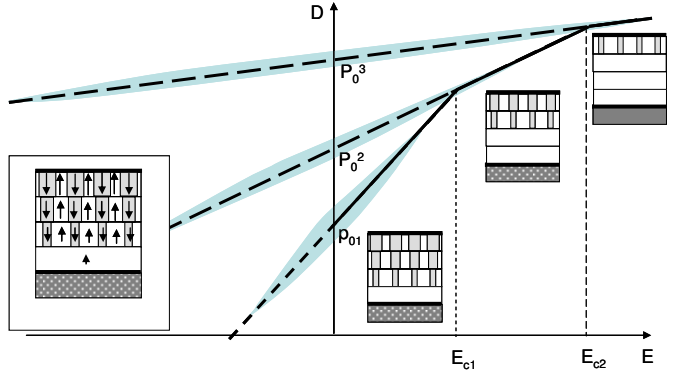

FIG.3 Polarization of a graded multilayer with four layers.

With increasing field layer 2 transforms to a single-domain state, then layer 3 transforms and so on, until all layers become single-domain ones. The change of polarization proceeds by movement of the boundary between a single-domain part and a polydomain part of the multilayer, i.e. by growth or shrinkage of wedge domains. This process can be described formally by the set of equations similar to Eq.(5) for single-domain layers and Eq.(4) for polydomain layers. The solutions of these equations for a multilayer containing one and two single-domain layers are described by Eq.(7) and (8) with $\alpha_1 + \alpha_2 < 1$. For the multilayer containing $i$ single-domain layers, the solution is:

$$D^{(i)} = \varepsilon^{(i)} E + P_r^{(i)}, \beta_k = 1/2(1 - D^{(i)}/p_{0k}) \text{ if } k > i,$$
$$\beta_k = 0 \text{ and if } k \leq i, \qquad (9)$$

where $\varepsilon^{(i)} = \left(\sum_{k=1}^{i} \frac{\alpha_k}{\varepsilon_k}\right)^{-1}$ is the dielectric susceptibility, and $P_r^{(i)} = \varepsilon^{(i)} \sum_{k=1}^{i} \frac{\alpha_k}{\varepsilon_k} p_{0k}$ is the remanent polarization of the multilayer at the field E, $E_{ci-1}<E<E_{ci}$, where $E_{ci-1} = (p_{0i} - P_r^{(i)})/\varepsilon^{(i-1)}$ (Fig.3). In each step the graded multilayer responds as "capacitors in series" which include only the layers in single-domain states. Since the stiffness of the polydomain layers close to zero, the external field is concentrated in the single-domain part. The energy of domain walls and non-uniform internal fields should decrease stability of a polydomain part of a graded multilayer and enhance its dielectric response. Therefore, at a low field the response of a polydomain part is controlled by change of polarization of a small single-domain part. Due to irreversibility of the transition from a polydomain to a single-domain state, partial hysteresis loops generated by AC field are observed in graded films (Fig.3) [2].

The linear approximation for the field (Eq.(5)) is used here for simplicity and can be avoided in calculations of P(E) for particular ferroelectrics. The presented theory can be easily expanded for continuously graded films if the change of their parameters ($P_0, \varepsilon$) is negligible along the characteristic lengths $l$ (1-10 nm). Otherwise, it is necessary to include gradient terms in the density of free energy.

The model of dense domain structure [17, 26] employed here does not describe the details of a domain morphology. Therefore, the predication of a wedge-like domain structure should be supported by the modeling and experimental studies. However, the presented model of a graded multilayer as a sequence of polydomain layers with variable fraction of domains gives the principle explanation of the peculiarities of dielectric behavior of graded films.

We are grateful to Dr. S.P.Alpay and Dr. J.V.Mantese for stimulating discussion. The support of NSF-DMR, grant# 0407517 is appreciated.